\documentstyle[aps,prb,preprint,tighten,eqsecnum,epsf,floats]{revtex}
\begin{document}
\draft
\title{Localization in a Disordered Multi-Mode
Waveguide with Absorption or Amplification}

\author{T.~Sh.~Misirpashaev,$^{a,b}$
J.~C.~J.~Paasschens,$^{a,c}$ and
C.~W.~J.~Beenakker,$^a$}
\address{$^a$Instituut-Lorentz, University of Leiden,
       P.O. Box 9506, 2300 RA Leiden, The Netherlands\\
$^b$Landau Institute for Theoretical Physics, 2 Kosygin
 Street, Moscow 117334, Russia\\
$^c$Philips Research Laboratories, 5656 AA Eindhoven, The
      Netherlands}
\maketitle
\begin{abstract}

An analytical and numerical study is presented of
transmission of radiation through a multi-mode waveguide
containing a random medium with
a complex dielectric constant~$\varepsilon=
\varepsilon'+i\varepsilon''$. Depending on the sign
of $\varepsilon''$, the medium is
absorbing or amplifying. The transmitted intensity decays
exponentially
$\propto\exp(-L/\xi)$ as the waveguide length $L\to\infty$,
regardless
of the sign of $\varepsilon''$. The localization length
$\xi$ is computed
as a function of the mean free path $l$, the absorption
or amplification
length $|\sigma|^{-1}$, and the number of modes in the
waveguide $N$. The
method used is an extension of the Fokker-Planck approach of
Dorokhov, Mello, Pereyra, and Kumar to non-unitary scattering
 matrices.
Asymptotically exact results are obtained for $N\gg1$ and
 $|\sigma|\gg1/N^2l$.
An approximate interpolation formula for all $\sigma$ agrees
 reasonably well
with numerical simulations.

\bigskip
\end{abstract}

\section{INTRODUCTION}

Localization of waves in one-dimensional random media has been
studied extensively, both for optical and for electronic systems.
\cite{Erd82,Pen94}
An analytical solution for the case of weak disorder
(mean free path $l$
much greater than the wavelength $\lambda$) was obtained as early
as 1959 by Gertsenshtein and Vasil'ev. \cite{Ger59}
The transmittance
$T$ (being the ratio of transmitted and incident intensity) has
a log-normal
distribution for large lengths $L$ of the system, with a mean
$\langle \ln T\rangle=-L/\xi$ characterized by a localization
length $\xi$ equal to the mean free path.

This early work was concerned with the propagation of classical
 waves, and hence included also the effect of absorption.
 In the presence of absorption
the transmittance decays faster, according to \cite{Ram87,Fre94}
$\langle\ln T\rangle=(\sigma-l^{-1})L$, where $|\sigma|$
is the inverse absorption
length ($\sigma<0$). Absorption is the result of a positive
imaginary part $\varepsilon''$ of the (relative) dielectric
constant $\varepsilon=\varepsilon'+i\varepsilon''$.
For a homogeneous $\varepsilon''$ one has
\begin{equation}
  \sigma = -2k \mathop{\rm{Im}}\sqrt{1+i\varepsilon''}
\approx-k\varepsilon''
\quad{\rm if\ }|\varepsilon''|\ll1,
  \label{sigdef}
\end{equation}
where $k$ is the wavenumber.
A negative $\varepsilon''$ corresponds to amplification by
stimulated emission of radiation, with
inverse amplification length $\sigma>0$.
Propagation of waves through amplifying one-dimensional
random media has been studied in
Refs.~\onlinecite{Pra94,Gup95,Zha95,Paa96,Fre96}.
In the limit $L\to\infty$ amplification also leads to a faster
decay of the transmittance,
according to $\langle \ln T\rangle=(-|\sigma|-l^{-1})L$.
\cite{Zha95,Paa96}

A natural extension of these studies is to waveguides which
contain more than a single propagating mode. Localization
in such ``quasi-one-dimensional'' systems has been studied
on the basis of a scaling theory, \cite{Tho77}
a supersymmetric field theory, \cite{Efe83} or a Fokker-Planck
equation. \cite{Dor82,Mel88b} It is found that the localization
length for $N$ modes is enhanced by a factor of order $N$
relative to the single-mode case.
These investigations were concerned with quantum mechanical,
rather than classical waves,
and therefore did not include absorption.
It is the purpose
of the present paper to extend the Fokker-Planck approach of
Dorokhov, Mello, Pereyra, and Kumar \cite{Dor82,Mel88b}
(DMPK) to include the effects on the
transmittance of a non-zero imaginary part of the dielectric
 constant.

According to the general duality relation of Ref. \onlinecite{Paa96},
the localization length is an even function of $\sigma$ for any $N$,
\begin{equation}
\xi(\sigma)=\xi(-\sigma).
\label{xiisxi}\end{equation}
 It follows
that both absorption and amplification lead to a faster decay of
the transmittance for large $L$. For $N\gg1$ we find that,
in good approximation,
\begin{equation}
{1\over\xi}={2\over (N+1)l}+(\sigma^2+2|\sigma|/l)^{1/2}.
\label{1.2}\end{equation}
This result becomes exact in the two limits
$|\sigma|\gg 1/N^2 l$ and
$|\sigma|\ll 1/N^2 l$. We compare with numerical solutions of the
Helmholtz equation, and find reasonably
good agreement over the whole range of $\sigma$.

The outline of the paper is as follows. In Sec.~II we formulate the
 scattering
problem and summarize the duality relation of
Ref.~\onlinecite{Paa96}. In Sec.~III
we derive a Fokker-Planck equation for the transmission
and reflection eigenvalues
${\cal T}_n$, ${\cal R}_n$, $n=1,2,\ldots N$.
These are eigenvalues of the matrix products
$tt^{\dagger}$ and $rr^{\dagger}$, respectively, where $t$ and $r$ are
the transmission and reflection matrices of the waveguide.
For $\sigma=0$ the Fokker-Planck
equation is the DMPK equation. \cite{Dor82,Mel88b}
A reduced Fokker-Planck equation, containing only the ${\cal R}_n$'s,
was previously obtained and studied
in Ref.~\onlinecite{Bee96}. To obtain the localization
length one needs to include
also the ${\cal T}_n$'s, which are no longer related to
the ${\cal R}_n$'s when $\sigma\neq0$.
We find that a closed Fokker-Planck equation containing
${\cal R}_n$'s and ${\cal T}_n$'s
exists only for $N=1$. If $N>1$ there appears an additional
set of ``slow variables,'' consisting of eigenvectors of $rr^{\dagger}$
in a basis where $tt^{\dagger}$
is diagonal. (These new variables do not appear when $\sigma=0$,
because then $rr^{\dagger}$ and $tt^{\dagger}$ commute.) Because of these
additional relevant variables we have not been able to make as
much progress in the solution of the Fokker-Planck equation
for $\sigma\neq0$ as one can for $\sigma=0$. \cite{RMP96}
In Sect.~IV we show that a closed evolution equation for
$\langle\ln T\rangle$ can be obtained if $|\sigma|\gg 1/N^2l$,
which leads to the second term in
Eq.~(\ref{1.2}). (This term could also have been obtained from the
incoherent radiative transfer theory for $\sigma<0$, but not
for $\sigma>0$.) To contrast
the multi-mode and single-mode cases, we also briefly discuss
in Sec.~IV the derivation of the localization length for $N=1$.
(Our $N=1$ results were
given without derivation in Ref.~\onlinecite{Paa96}.) Finally,
in Sec.~V we compare the analytical results for the multi-mode
case with numerical simulations.

\section{FORMULATION OF THE SCATTERING PROBLEM}

We consider a random medium
of length $L$ with a spatially fluctuating dielectric constant
$\varepsilon=\varepsilon'+i\varepsilon''$,
embedded in an $N$-mode waveguide with $\varepsilon=1$.
The scattering matrix
$S$ is a $2N\times2N$ matrix relating incoming and outgoing modes
at some frequency $\omega$. It
has the block structure
\begin{equation}
  S=\pmatrix{r' & t'\cr t & r},
\end{equation}
where $t$, $t'$ are the transmission matrices and $r$, $r'$ the
reflection
 matrices. We introduce the sets of transmission and reflection
 eigenvalues
$\{{\cal T}^{\vphantom1}_n\}$,
$\{{\cal T}'_n\}$, $\{{\cal R}^{\vphantom1}_n\}$, $\{{\cal R}'_n\}$,
being the
eigenvalues of, respectively, $tt^\dagger$,
$t't'^\dagger$,
$rr^\dagger$,
$r'r'^\dagger$.
Total transmittances and reflectances are defined as
\begin{mathletters}
\label{TRboth}
\begin{eqnarray}
& T=N^{-1}\mathop{\rm Tr} tt^{\dagger},&\qquad R=N^{-1}\mathop{\rm Tr}
rr^{\dagger},\label{TR}\\
&  T'=N^{-1}\mathop{\rm Tr} t't'{}^{\dagger},
&\qquad R'=N^{-1}\mathop{\rm Tr}
r'r'{}^{\dagger}.\label{TRp}
\end{eqnarray}
\end{mathletters}%
Here $T$ and $R'$ are the transmitted and reflected intensity divided
by the
incident intensity from the left. Similarly, $T'$ and $R$ correspond
to incident intensity from the right. By taking the trace in
Eq.~(\ref{TRboth})
we are assuming diffuse illumination, i.e.\ that the incident intensity
is
equally distributed over the $N$ modes. Two systems which differ only
in
the sign of $\varepsilon''(\vec r)$ are called dual. Scattering
matrices of
dual systems are related by \cite{Paa96}
\begin{equation}
  S(\varepsilon'')S^{\dagger}(-\varepsilon'')=1.
\label{GF}\end{equation}
This duality relation takes the place of the unitarity constraint  when
$\varepsilon''\neq0$.

An optical system usually possesses time-reversal symmetry, as a
result of which $S(\varepsilon'')S^*(-\varepsilon'')=1$. Combining this
relation with Eq.~(\ref{GF}), we find that
$S=S^{\rm T}$ is a symmetric matrix. Hence
${\cal T}_n={\cal T}_n'$ and $T=T'$. (The reflectances $R$ and $R'$ may
differ.)
The case of broken time-reversal symmetry
might also be physically relevant, \cite{Rik96} and is included here
for completeness. In the absence of time-reversal symmetry $S$ is
an arbitrary complex matrix.

The duality relation
(\ref{GF}) has consequences for the
reflection and transmission eigenvalues of two dual systems.
\cite{Paa96}
If $N=1$ the relation
\begin{equation}
T(\varepsilon'')/R(\varepsilon'')=T'(-\varepsilon'')/R'(-\varepsilon'')
\label{ToverR}\end{equation}
holds for all $L$. If $N\geq1$ we have two relations for $L\to\infty$,
\begin{eqnarray}
\lim_{L\to\infty}{\cal R}_n(\varepsilon'')&=&\lim_{L\to\infty}{\cal
R}_n^{-1}(-\varepsilon''),
\label{RDuality}\\
\lim_{L\to\infty}L^{-1}\ln{\cal T}_n(\varepsilon'')&=&
\lim_{L\to\infty}L^{-1}\ln{\cal T}_n(-\varepsilon'').
\end{eqnarray}
The transmittance $T=N^{-1}\sum_n{\cal T}_n$
is dominated by the largest transmission eigenvalue,
hence
\begin{equation}
\lim_{L\to\infty}L^{-1}\ln T(\varepsilon'')=
\lim_{L\to\infty}L^{-1}\ln T(-\varepsilon'').
\end{equation}
In other words, two dual systems have the same localization length,
as stated in Eq.~(\ref{xiisxi}).

\section{FOKKER-PLANCK EQUATION}
We derive a Fokker-Planck equation for the evolution of the
distribution
of scattering matrices with increasing length $L$ of the waveguide. In
the
absence of gain or loss ($\sigma=0$), the evolution equation is known
as the
Dorokhov-Mello-Pereyra-Kumar (DMPK) equation. \cite{Dor82,Mel88b}
Original derivations
of this equation relied on the unitarity of the
scattering matrix, making use of the invariant measure
on the unitary group and the polar
decomposition of a unitary matrix.
These derivations cannot readily be generalized  to the case
$\sigma\neq0$, in particular because the scattering matrix
no longer admits a polar decomposition.
(This means that the matrix products $rr^{\dagger}$ and $tt^{\dagger}$ do not
commute.)
The alternative derivation of the DMPK equation of
Ref.~\onlinecite{RMP96} does not use the polar decomposition and is
suitable for our purpose.

Without loss of generality we can write the transmission and reflection
submatrices of the scattering matrix as follows,
\begin{equation}
S=\pmatrix{r' & t' \cr t & r}=
\pmatrix{U\sqrt{{\bf R}'}W &
U'\sqrt{{\bf T}'}Z \cr
V\sqrt{{\bf T}}W' & -V'\sqrt{{\bf R}}Z'}.
\label{SMatPar}\end{equation}
Here $U$, $U'$, $V$, $V'$, $W$, $W'$, $Z$, $Z'$ are $N\times N$ unitary
matrices, while ${\bf R}$, ${\bf R}'$, ${\bf T}$, ${\bf
T}'$
are diagonal matrices whose elements are
the reflection and transmission eigenvalues
$\{{\cal R}_n\}$, $\{{\cal R}'_n\}$,
$\{{\cal T}_n\}$, $\{{\cal T}'_n\}$.
For $\sigma=0$, the unitarity constraint $SS^{\dagger}=1$ implies
$U=U'$, $V=V'$, $W=W'$, $Z=Z'$,
and ${\bf R}={\bf R}'=1-{\bf T}=1-{\bf T}'$.
Eq.~(\ref{SMatPar}) then constitutes
the polar decomposition of the scattering matrix. In this case one can
derive a Fokker-Planck equation for the evolution of only transmission
or only reflection
eigenvalues.
If $\sigma\neq0$, the Fokker-Planck equation contains both the
transmission and
reflection eigenvalues, as well as elements of
the matrix  $Q=V^{\dagger} V'$
relating eigenvectors of $tt^{\dagger}$ and
$rr^{\dagger}$. The only constraint on the scattering matrix if
$\sigma\neq0$
is imposed by time-reversal symmetry, which requires
$S=S^{\rm T}$, hence $W=U^{\rm T}$, $Z=V^{\rm T}$, $W'=U'{}^{\rm T}$,
$Z'=V'{}^{\rm T}$, ${\bf T}={\bf T}'$.

The Fokker-Planck equation describes the evolution of slow variables
after the elimination of fast variables. In our problem fast variables
vary on the scale of the wavelength $\lambda$, while slow variables
vary on the scale of the mean free path $l$ or the amplification length
$|\sigma|^{-1}$. We assume that both $l$ and $|\sigma|^{-1}$ are much
greater
than $\lambda$. (This requires $|\varepsilon''|\ll1$.) The slow
variables
include $\{{\cal R}_n\}$, $\{{\cal T}_n\}$ and
elements of $Q=V^{\dagger} V'$. We denote
this set of slow variables collectively
by $\{\Phi_n\}$. Each $\Phi_i$ is incremented by
$\delta\Phi_i$ if a thin slice of length $\delta L$ ($\lambda\ll\delta
L\ll l$)
is added to the waveguide of length $L$. The increments
are of order $(\delta L/l)^{1/2}$ and can be calculated
perturbatively. We specify an appropriate statistical ensemble
for the scattering matrix $\delta S$ of the thin slice and
compute moments of $\delta\Phi_i$.
 The first two moments are of order $\delta L/l$,
\begin{mathletters}\label{PhiMoments}\begin{eqnarray}
&\langle\delta\Phi_i\rangle&=a_i\,\delta L/l+{\cal O}(\delta
L/l)^{3/2},\\
&\langle\delta\Phi_i\delta\Phi_j\rangle&=a_{ij}\,\delta L/l +
{\cal O}(\delta L/l)^{3/2}.
\end{eqnarray}\end{mathletters}%
Higher moments have no term of order $\delta L/l$. According to the
general
theory of Brownian motion, \cite{Kam81}
the Fokker-Planck equation for the joint probability distribution
$P(\{\Phi_n\},L)$ reads
\begin{equation}
l{\partial P\over\partial L}=-\sum_i{\partial\over\partial\Phi_i}
 a_i P+{1\over2}\sum_{ij}{\partial^2\over\partial\Phi_i\partial\Phi_j}
 a_{ij} P.
\label{FPEGeneral}\end{equation}

The average $\langle\cdots\rangle$ in Eq.~(\ref{PhiMoments}) is defined
by
the statistics of $\delta S$. We specify this statistics using
simplifying features of the waveguide geometry (length ${\gg}$ width),
which justify the equivalent-channel or isotropy approximation.
\cite{Mel88b,Mel92}
 We assume
that amplification or absorption in the thin slice is independent
of the scattering channel. This entails the relation
\begin{equation}
\delta S\delta S^{\dagger}=1+\bar\sigma\,\delta L,
\label{InfFlux}\end{equation}
where $\bar\sigma $
is a modal and spatial average
of the inverse amplification length $\sigma$. If $\varepsilon''$ is
spatially constant, one has
\begin{equation}
\bar\sigma=-{2k\over N}\sum_{n=1}^N{\rm Im}\,(1-\omega_n^2/\omega^2+
i\varepsilon'')^{1/2},
\end{equation}
where $\omega_n$ is the cutoff frequency of mode $n$.
For $N\to\infty$, the sum over modes
can be replaced by an integral. The result depends on the
dimensionality
of the waveguide,
\begin{mathletters}\label{sigmabar}\begin{eqnarray}
\bar\sigma&=&-2k\varepsilon'', {\rm\ for\ a\ 3D\ waveguide},\\
\bar\sigma&=&-(\pi/2)k\varepsilon'', {\rm\ for\ a\ 2D\ waveguide},
\end{eqnarray}\end{mathletters}%
where we have used that $|\varepsilon''|\ll1$.

Eq.~(\ref{InfFlux}) ensures the
existence of a polar decomposition for $\delta S$,
\begin{equation}
\delta S=
\pmatrix{U_0\sqrt{\delta{\bf R}}W_0 &
U_0\sqrt{\delta{\bf T}}Z_0 \cr
V_0\sqrt{\delta{\bf T}}W_0 & -V_0\sqrt{\delta{\bf R}}Z_0},
\label{polar}\end{equation}
with $\delta{\bf T}+\delta{\bf R}=1+\bar\sigma\delta L$.
Note that a polar decomposition for $\delta S$ does not imply a polar
decomposition for $S$, because the special block structure of
Eq.~(\ref{polar}) is lost upon composition of scattering matrices.
We make the isotropy
assumption that the matrices $U_0$, $V_0$, $W_0$, $Z_0$ are uniformly
distributed in the unitary group.
In the presence of time-reversal symmetry
 one has $W_0^{\vphantom{T}}=U_0^{\rm T}$
and $Z_0^{\vphantom{T}}=V_0^{\rm T}$. In the absence
of time-reversal symmetry  all four unitary matrices are
independent. The diagonal matrices
$\delta{\bf R}$ and $\delta{\bf T}$
may have arbitrary distributions. We specify the first moments,
\begin{mathletters}\label{SliceDiag}\begin{eqnarray}
&&\langle\mathop{\rm Tr} \delta{\bf R}\rangle=N\,\delta L/l, \\
&&\langle\mathop{\rm Tr} \delta{\bf T}\rangle=N+N(\gamma-1)\,\delta
L/l,
\end{eqnarray}\end{mathletters}%
where we have defined $\gamma=\bar\sigma l$.
The mean free path $l$ in Eq.~(\ref{SliceDiag}) is related to the mean
free path $l_{\rm tr}$ of radiative transfer theory by \cite{RMP96}
\begin{mathletters}\label{lnotr}\begin{eqnarray}
l&=&(4/3)l_{\rm tr}, {\rm\ for\ a\ 3D\ waveguide},\\
l&=&(\pi/2)l_{\rm tr}, {\rm\ for\ a\ 2D\ waveguide}.
\end{eqnarray}\end{mathletters}%
This completes the specification of the statistical ensemble for
$\delta S$.

We need the increments $\Delta{\cal R}_n$, $\Delta{\cal T}_n$
of reflection and transmission eigenvalues
to first order in $\delta L/l$,
\begin{mathletters}\begin{eqnarray}
\Delta{\cal R}_n&=&\Delta R^{(1)}_{nn}+\Delta R^{(2)}_{nn}+\sum_{m\neq
n}
{\Delta R^{(1)}_{nm}\Delta R^{(1)}_{mn}\over {\cal R}_n-{\cal R}_m},\\
\Delta{\cal T}_n&=&\Delta T^{(1)}_{nn}+\Delta T^{(2)}_{nn}+\sum_{m\neq
n}
{\Delta T^{(1)}_{nm}\Delta T^{(1)}_{mn}\over {\cal T}_n-{\cal T}_m}.
\end{eqnarray}\end{mathletters}%
The matrices of perturbation $\Delta R^{(1)}$, $\Delta R^{(2)}$,
$\Delta T^{(1)}$, $\Delta T^{(2)}$ are expressed through unitary
matrices $Q=V^{\dagger} V'$, $\tilde U=Z'U_0$, $\tilde W=W_0V'$
 and diagonal
matrices $\bf T$, $\bf R$, $\delta{\bf T}$,
$\delta{\bf R}$,
\begin{mathletters}\begin{eqnarray}
\Delta R^{(1)}&=&\left[\sqrt{\bf R}\tilde U\sqrt{\delta\bf
R}\tilde W(1-{\bf R})+
{\rm H.c.}\right],\\
\Delta R^{(2)}&=&-\sqrt{{\bf R}}\tilde U(1-\delta{\bf T})\tilde
U^{\dagger}\sqrt{{\bf R}}+
\tilde W^{\dagger}\delta{\bf R}\tilde W+
\sqrt{{\bf R}}\tilde U\sqrt{\delta {\bf R}}\tilde W{\bf R}
\tilde W^{\dagger}\sqrt{\delta{\bf R}}\tilde U^{\dagger}
\sqrt{{\bf R}} \cr
&&-\left[\case{1}{2}\tilde W^{\dagger}(1-\delta{\bf T})\tilde W{\bf
R}+
\sqrt{{\bf R}}\tilde U\sqrt{\delta{\bf R}}\tilde
W\sqrt{{\bf R}}\tilde U\sqrt{\delta{\bf R}}\tilde W
(1-{\bf R})+{\rm H.c.}\right],\\
\Delta T^{(1)}&=&-\left[Q\sqrt{{\bf R}}\tilde U\sqrt{\delta
{\bf R}}\tilde W Q^{\dagger}{\bf T}+{\rm H.c.}
\right],\\
\Delta T^{(2)}&=&Q\sqrt{{\bf R}}\tilde U\sqrt{\delta {\bf
R}}\tilde W Q^{\dagger}{\bf T} Q\tilde W^{\dagger}
\sqrt{\delta{\bf R}}\tilde U^{\dagger}\sqrt{{\bf R}}Q^{\dagger}\cr
&&-\left[\case{1}{2}Q\tilde W^{\dagger}
(1-\delta{\bf T})\tilde W Q^{\dagger}
{\bf T}-
Q\sqrt{{\bf R}}\tilde U\sqrt{\delta {\bf R}}\tilde
W\sqrt{{\bf R}}\tilde U\sqrt{\delta{\bf R}}
\tilde W Q^{\dagger} {\bf T} +{\rm H.c.}\right].
 \end{eqnarray}\end{mathletters}%
(The abbreviation H.c. stands for Hermitian conjugate.)
The moments (\ref{PhiMoments}) are computed by first
averaging over the unitary  matrices
$U_0$, $W_0$ and then averaging over $\delta{\bf R}$ and
$\delta{\bf T}$
using Eq.~(\ref{SliceDiag}). Averages over unitary matrices follow from
\begin{mathletters}\begin{eqnarray}
 \langle U^{\vphantom*}_{nk}U^*_{ml}\rangle&=&{1\over
 N}\delta_{nm}\delta_{kl},\\
\langle U^{\vphantom*}_{nk} U^{\vphantom*}_{mk} U^*_{pl} U^*_{ql}
\rangle&=&{1\over N(N+1)}
\delta_{kl}\left(\delta_{np}\delta_{mq}+\delta_{nq}\delta_{mp}\right).
\end{eqnarray}\end{mathletters}%
Without time-reversal symmetry averages over $U_0$ and $W_0$ are
independent. With time-reversal symmetry we have
$W^{\vphantom T}_0=U_0^{\rm T}$ so that only a single average remains.
The results are

\medskip

\centerline{\em With time-reversal symmetry}

\begin{mathletters}\label{Momentsb1}\begin{eqnarray}
(l/\delta L)\langle\delta{\cal R}_n\rangle&=&
1+2(\gamma-1){\cal R}_n+{{\cal R}_n\over N+1}\Bigl({\cal
R}_n+\sum_m{\cal R}_m\Bigr)\cr
&&\mbox{}+{1\over N+1}\sum_{m\neq n}{{\cal R}_n(1-{\cal R}_m)^2+{\cal
R}_m(1-{\cal R}_n)^2
\over {\cal R}_n-{\cal R}_m},
\label{deltaRb1}\\
(l/\delta L)\langle\delta{\cal R}_n\delta{\cal
R}_m\rangle&=&{4\delta_{nm}\over N+1}{\cal R}_n
(1-{\cal R}_n)^2,\label{deltaR2b1}\\
(l/\delta L)\langle\delta{\cal T}_n\rangle&=&{\cal T}_n(\gamma-1)+
{{\cal T}_n\over N+1}\Bigl(A_{nn}+\sum_{m\neq n}{{\cal T}_mA_{nn}+{\cal
T}_nA_{mm}\over
{\cal T}_n-{\cal T}_m}\cr
&&\mbox{}+F_{nn}+\sum_{m\neq n}F_{nm}{{\cal T}_n+{\cal T}_m\over {\cal
T}_n-{\cal T}_m}\Bigr), \label{deltaTb1}\\
(l/\delta L)\langle\delta{\cal T}_n\delta{\cal T}_m\rangle&=&
{2\over N+1}\left(\delta_{nm}{\cal T}_n^2A_{nn}+{\cal T}_n{\cal T}_m
F_{nm}\right),\label{deltaT2b1}\\
(l/\delta L)\langle\delta{\cal T}_n\delta{\cal R}_m\rangle&=&-
{4\over N+1}{\cal T}_n{\cal R}_m(1-{\cal R}_m)|Q_{nm}|^2.
\label{deltaTRb1}
\end{eqnarray}\end{mathletters}%

\centerline{\em Without time-reversal symmetry}

\begin{mathletters}\label{Momentsb2}\begin{eqnarray}
(l/\delta L)\langle\delta{\cal R}_n\rangle&=&
1+2(\gamma-1){\cal R}_n+{{\cal R}_n\over N}\sum_m{\cal R}_m\cr
&&\mbox{}+{1\over N}\sum_{m\neq n}{{\cal R}_n(1-{\cal R}_m)^2+{\cal
R}_m(1-{\cal R}_n)^2
\over {\cal R}_n-{\cal R}_m},
\label{deltaR}\\
(l/\delta L)\langle\delta{\cal R}_n\delta{\cal R}_m\rangle&=&
{2\delta_{nm}\over N}{\cal R}_n
(1-{\cal R}_n)^2,\label{deltaR2}\\
(l/\delta L)\langle\delta{\cal T}_n\rangle&=&{\cal T}_n(\gamma-1)+
{{\cal T}_n\over N}\Bigl(A_{nn}+\sum_{m\neq n}{{\cal T}_mA_{nn}+{\cal
T}_nA_{mm}\over
{\cal T}_n-{\cal T}_m}\Bigr), \label{deltaT}\\
(l/\delta L)\langle\delta{\cal T}_n\delta{\cal T}_m\rangle&=&
{2\delta_{nm}\over N}{\cal T}_n^2A_{nn},
\label{deltaT2}\\
(l/\delta L)\langle\delta{\cal T}_n\delta{\cal R}_m\rangle&=&-
{2\over N}{\cal T}_n{\cal R}_m(1-{\cal R}_m)|Q_{nm}|^2. \label{deltaTR}
\end{eqnarray}\end{mathletters}%
We have abbreviated $A_{nm}=(Q{\bf R} Q^{\dagger})_{nm}$ and
$F_{nm}=|(Q\sqrt{{\bf R}}Q^{\rm T})_{mn}|^2$.

The moments of $\delta{\cal R}_n$ contain only the set of reflection
eigenvalues
$\{{\cal R}_n\}$, so that from Eq.~(\ref{FPEGeneral})
we can immediately write down a Fokker-Planck equation for the
distribution
of the ${\cal R}_n$'s.
In terms of variables $\mu_n=1/({\cal R}_n{-}1)\in
(-\infty,-1)\cup(0,\infty)$
 it reads \cite{Bee96}
\begin{eqnarray}
l{\partial\over\partial L}P(\{\mu_n\},L)&=&{2\over\beta
N{+}2{-}\beta}\sum_{n=1}^N{\partial\over\partial\mu_n}
\mu_n(1+\mu_n)\cr
&&\mbox{}\times\left[{\partial P\over\partial\mu_n}+\beta P\sum_{m\neq
n}
{1\over \mu_m-\mu_n}+\gamma (\beta N{+}2{-}\beta)P\right],
\label{ManyRFPE}\end{eqnarray}
where the symmetry index $\beta=1(2)$ corresponds to the case of
unbroken (broken) time-reversal symmetry.
The evolution of the reflection eigenvalues is independent
of the transmission eigenvalues---but not vice versa. The evolution of
the ${\cal T}_n$'s depends on the ${\cal R}_n$'s, and in addition on
the
slow variables contained in the unitary matrix $Q$. To obtain a closed
Fokker-Planck equation we also need to compute increments and moments
of $Q$. The resulting expressions are lengthy and will not be written
down
here.

In the single-mode case ($N=1$) this complication does not arise,
because $Q=e^{i\phi}$ drops out of the scalars $A$ and $F$.
The single transmission and reflection eigenvalues ${\cal T}$, ${\cal
R}$
coincide with the transmittance and reflectance $T$, $R$ defined by
Eq.~(\ref{TRboth}).
The resulting Fokker-Planck equation is \cite{Paa96}
\begin{eqnarray}
l{\partial P\over\partial L}=-&&{\partial\over\partial
R}\left[(1-R)^2+2\gamma R\right]P+
{\partial^2\over\partial R^2}R(1-R)^2 P\cr
&&-{\partial\over\partial T}T(\gamma-1+R)P+{\partial^2\over\partial
T^2}T^2RP-2{\partial^2\over\partial T\partial R}TR(1-R)P.
\label{FPE_Plain}
\end{eqnarray}
In the case of absorption ($\gamma<0$), Eq.~(\ref{FPE_Plain}) is
equivalent to
the moments equations of Ref.~\onlinecite{Fre94}.

\section{Localization length}
The limit $L\to\infty$ of the distribution of the reflection
eigenvalues
follows directly from Eq.~(\ref{ManyRFPE}), by equating the
left-hand-side
to zero.
The resulting
distribution $P_\infty$ is that of the Laguerre ensemble of random
matrix
theory, \cite{Bee96}
\begin{equation}
  P(\{\mu_n\})\propto\prod\limits_{i<j}|\mu_j-\mu_i|^\beta
  \prod\limits_k \exp[-\gamma(\beta N{+}2{-}\beta)\mu_k].
  \label{ALaguerre}
\end{equation}
The distribution looks the same for both signs of $\gamma$, but
the support (and the normalization constant) is different:
$\mu_n>0$ for $\gamma>0$, and $\mu_n<-1$ for $\gamma<0$.
To determine the localization length we need the distribution
of the transmission eigenvalues in the large-$L$ limit. We consider
the cases $N=1$ and $N\gg1$.
\subsection{Single-mode waveguide}
We compute the distribution $P(T,L)$ of the transmission probability
through a single-mode waveguide in the limit $L\to\infty$. In the case
of absorption ($\gamma<0$) this calculation was done by Rammal
and Doucot, \cite{Ram87} and by Freilikher, Pustilnik, and Yurkevich.
\cite{Fre94} We generalize their results to the case of amplification
($\gamma>0$). The two cases are essentially different because,
while the mean value of $R$ is finite in the case of absorption,
\begin{equation}
\langle R\rangle_\infty=1-2\gamma e^{-2\gamma}\,{\rm
Ei}(2\gamma),\qquad
{\rm for\ }    \gamma<0,
\label{RMeanAb}\end{equation}
it diverges in the case of amplification.
The mean value of $\ln R$ is finite in both cases,
\begin{equation}
\langle\ln R\rangle_\infty=\cases{
{\bf C}+\ln 2\gamma-e^{2\gamma}\,{\rm Ei}(-2\gamma),& for
$\gamma>0$,\cr
-{\bf C}-\ln(-2\gamma) +e^{-2\gamma}\,{\rm Ei}(2\gamma),& for
$\gamma<0$.\cr}
\label{lnRAv}\end{equation}
Here $\bf C$ is Euler's constant and
${\rm Ei}(x)=\int_{-\infty}^x dt\,e^t/ t$
is the exponential integral.
The relation
\begin{equation}
\langle\ln R(\gamma)\rangle_\infty=-\langle\ln R(-\gamma)\rangle_\infty
\label{RAvDual}\end{equation}
holds, in accordance with the duality relation (\ref{RDuality}).

We now show that the asymptotic $L\to\infty$ distribution of $T$ is
log-normal,
with mean and variance of $\ln T$ given by
\begin{mathletters}\label{lTMean}
\begin{eqnarray}
  \langle\ln  T\rangle&=&-(1+|\gamma|)\,L/l+2c(\gamma)+{\cal
  O}(l/L),\label{lT}\\
  c(\gamma)&=&\cases{
    0,                                                     & for
$\gamma<0$, \cr
      {\bf C}+\ln  2\gamma-e^{2\gamma}\,{\rm Ei}(-2\gamma), & for
$\gamma>0$,}
  \label{lTint}
\end{eqnarray}\end{mathletters}%
\begin{equation}
  \mathop{\rm{var}}\ln  T= \Bigl[2+4|\gamma| e^{2|\gamma|}{\rm
Ei}(-2|\gamma|)\Bigr]
  \, L/l +{\cal O}(1).
  \label{lTVar}
\end{equation}
The constant $c(\gamma)\approx -2\gamma\ln\gamma$ if $0<\gamma\ll1$.
Note that $\mathop{\rm{var}}\ln  T\ll\langle\ln  T\rangle^2$ for
$L/l\gg1$.
The localization length
$\xi=l\,(1+|\gamma|)^{-1}$ is independent of the sign of $\gamma$, in
accordance
with the duality relation~(\ref{xiisxi}).

These results are easy to establish for the case of absorption,
when  Eq.~(\ref{FPE_Plain}) implies the evolution equations
\cite{Ram87,Fre94}
\begin{equation}
l{\partial\over\partial L}\langle\ln T\rangle=-1+\gamma,
\qquad
l{\partial\over\partial L}\mathop{\rm{var}}\ln T=2\langle R\rangle,
\qquad {\rm for\ } \gamma<0.
\label{evoeq}\end{equation}
Making use of the initial condition $T\to1$ for $L\to0$ and the
asymptotic value (\ref{RMeanAb}) of $\langle R\rangle$, one readily
obtains Eqs.~(\ref{lTMean})
and (\ref{lTVar}) for $\gamma<0$.

In the case of amplification, the evolution equations (\ref{evoeq})
hold
only for lengths $L$ smaller than
 $L_c\simeq l\,c(\gamma)/|\gamma|$.
For $L\lesssim L_c$
stimulated emission {\em enhances} transmission through the
waveguide.
On larger length scales stimulated emission {\em reduces}
transmission. Technically, the evolution equations (\ref{evoeq}) break
down for $L\to\infty$ because the integration by parts of the
Fokker-Planck
equation produces a non-zero boundary term if $L>L_c$. To extend
Eqs.~(\ref{lTMean})
and (\ref{lTVar})
to the case $\gamma>0$ we use the duality relation (\ref{ToverR}).
It implies that for $N=1$
the distribution of the ratio $T/R$ is an even function of $\gamma$.
Eq.~(\ref{lTMean}) for $\gamma>0$ follows directly from the equality
\begin{equation}
\langle\ln T(\gamma)/R(\gamma)\rangle=\langle\ln
T(-\gamma)/R(-\gamma)\rangle,
\end{equation}
which holds for all $L$, plus Eq.~(\ref{RAvDual}), which holds
for $L\to\infty$.
The constant $c(\gamma)$
for $\gamma>0$ equals $\langle\ln R(\gamma)\rangle_\infty$ and is
substituted from Eq.~(\ref{lnRAv}).
The duality of $T(\gamma)/R(\gamma)$ also implies Eq.~(\ref{lTVar}) for
the
variance, provided the
covariance $\langle\langle\ln T\ln R\rangle\rangle=
\langle \ln T\ln R\rangle-
\langle\ln T\rangle\langle\ln R\rangle$ remains finite
as $L\to\infty$. We have checked this directly
from the Fokker-Planck equation
(\ref{FPE_Plain}), and found the finite large-$L$ limit
\begin{eqnarray}
\langle\langle\ln T\ln R\rangle\rangle_\infty&=&
-2e^{2\gamma}\,{\rm Ei}(-2\gamma)c(\gamma)-c(\gamma)^2\cr
&&\mbox{}-2\gamma\int_0^\infty d\mu\,  e^{-2\gamma\mu}
\left[\ln^2(1+\mu)-\ln^2\mu\right], \quad {\rm for\ } \gamma>0.
\end{eqnarray}

\subsection{Multi-mode waveguide}
We next consider a waveguide with $N\gg1$ modes.
We compute the localization length
$\xi=-\lim_{L\to\infty}L^{-1}\langle\ln T\rangle$
in the case of absorption, and include the case of
amplification invoking
duality.
For absorption the average reflectance
$\langle R\rangle=N^{-1}\langle\sum_k(1+1/\mu_k)\rangle$
remains finite as $L\to\infty$. The large-$L$
limit $\langle R\rangle_\infty$ follows from the distribution
(\ref{ALaguerre}),
using known formulas for the eigenvalue density in the Laguerre
ensemble.
\cite{Nag93} For $|\gamma|N^2\gg1$ the result is
\begin{equation}
\langle R\rangle_\infty=1+|\gamma|-\sqrt{|\gamma|(2+|\gamma|)}+{\cal
O}(1/N), \qquad    \gamma<0.
\label{RManyMean}\end{equation}

The evolution of transmission eigenvalues is governed by the
Fokker-Planck
equation (\ref{FPEGeneral}), with coefficients given by
(\ref{PhiMoments}),
(\ref{Momentsb1}), and (\ref{Momentsb2}). Each ${\cal T}_n$ has its own
localization
length $\xi_n=-\lim_{L\to\infty}L^{-1}\ln{\cal T}_n$. We order the
$\xi_n$'s from
large to small,  $\xi_1>\xi_2>\dots>\xi_N$. This implies that for
$L\to\infty$
the separation of the ${\cal T}_n$'s becomes exponentially large,
${\cal T}_1\gg{\cal T}_2\gg\dots\gg{\cal T}_N$. Hence we may
approximate
\begin{mathletters}\begin{eqnarray}
{{\cal T}_n+{\cal T}_m\over{\cal T}_n-{\cal T}_m}&\approx&\cases{
-1, & for $n>m$, \cr
1, & for $n<m$,}\\
{{\cal T}_n A_{mm}+{\cal T}_m A_{nn}\over{\cal T}_n-{\cal
T}_m}&\approx&\cases{
-A_{nn}, & for $n>m$,\cr
A_{mm}, & for $n<m$.}
\end{eqnarray}\end{mathletters}%
The Fokker-Plank equation  (\ref{FPEGeneral}) simplifies considerably
and
leads to the following equation for the largest transmission
eigenvalue:
\begin{equation}
l{\partial\over\partial L}\langle\ln{\cal T}_1\rangle=
\cases{
-1-|\gamma|+\langle R\rangle-{1\over N+1}\langle A_{11}+F_{11}\rangle,
& for $\beta=1$,\cr
-1-|\gamma|+\langle R\rangle-{1\over N}\langle A_{11}\rangle, & for
$\beta=2$.}
\label{4.13}\end{equation}

For $|\gamma|N^2\gg1$ we may substitute Eq.~(\ref{RManyMean}) for
$\langle R\rangle$ and omit the terms with $\langle A_{11}\rangle$ and
 $\langle F_{11}\rangle$. The resulting localization length is given by
\begin{equation}
l/\xi=\sqrt{|\gamma|(2+|\gamma|)}+{\cal O}(1/N).
\label{4.14}\end{equation}
Because of duality, Eq.~(\ref{4.14}) holds regardless of the sign of
$\gamma$.
It agrees with radiative transfer theory for $\gamma<0$, but not for
$\gamma>0$.
Indeed, the exponential decay of the transmitted intensity in the case
of
amplification is  an interference effect, which is not contained
in the theory of radiative transfer.

Eq.~(\ref{4.14}) is asymptotically exact for $|\gamma|\gg1/N^2$. For
smaller
$|\gamma|$ we cannot compute $\xi$ rigorously because the distribution
of the matrices
${\bf A}$ and ${\bf F}$ is not known.
An interpolative formula for all $\gamma$
can be obtained by substituting
for $\langle A_{11}\rangle$ and
$\langle F_{11}\rangle$ in Eq.~(\ref{4.13})
 their $L\to\infty$ limits when $\gamma=0$, which are
$\langle A_{11}\rangle=\langle F_{11}\rangle=1$.
In this way, we arrive at the localization length
\begin{equation}
\xi=l\left[{2\over\beta
N{+}2{-}\beta}+\sqrt{|\gamma|(2+|\gamma|)}\right]^{-1},
\label{ximax}\end{equation}
which interpolates between the known \cite{Efe83,Dor82,MPS} value of
$\xi$ for $\gamma=0$
and Eq.~(\ref{4.14}) for $|\gamma|\gg1/N^2$.

The localization length $\xi$ is the largest of the
eigenvalue-dependent
localization lengths $\xi_n$. What about the other $\xi_n$'s? For
$\gamma=0$ it is known \cite{Dor82,Mel88b,MPS} that the inverse
localization lengths are equally spaced, and satisfy  the sum rule
$\sum_n1/\xi_n=N/l$. We have not succeeded in deriving the spacings
for $\gamma\neq0$, but we have been able to derive the sum rule
from the Fokker-Planck equation (by computing the $L$-dependence
of $\langle\sum_n\ln{\cal T}_n\rangle$). The result is exact and reads
\begin{equation}
l\sum_{n=1}^N\xi_n^{-1}=(1+|\gamma|) N.
\end{equation}

\section{NUMERICAL RESULTS}\label{SectionNumerics}
To test the analytical predictions on a model system,
 we have numerically solved a discretized
version of the Helmholtz equation,
\begin{equation}
  \left[\nabla^2+ k^2\varepsilon(\vec r)\right]E(\vec r)=0,
\label{Helm}\end{equation}
on a two-dimensional
square lattice (lattice constant $d$, length $L$, width $W$).
The real part $\varepsilon'$ of the dielectric constant was chosen
randomly from site to site with a uniform distribution between
$1\pm\delta\varepsilon$.
The imaginary part $\varepsilon''$ was the same at all sites.
 The scattering matrix
was computed using the recursive Green function technique. \cite{Bar91}

\begin{figure}[ht]
\epsfxsize=0.9\hsize
\epsffile{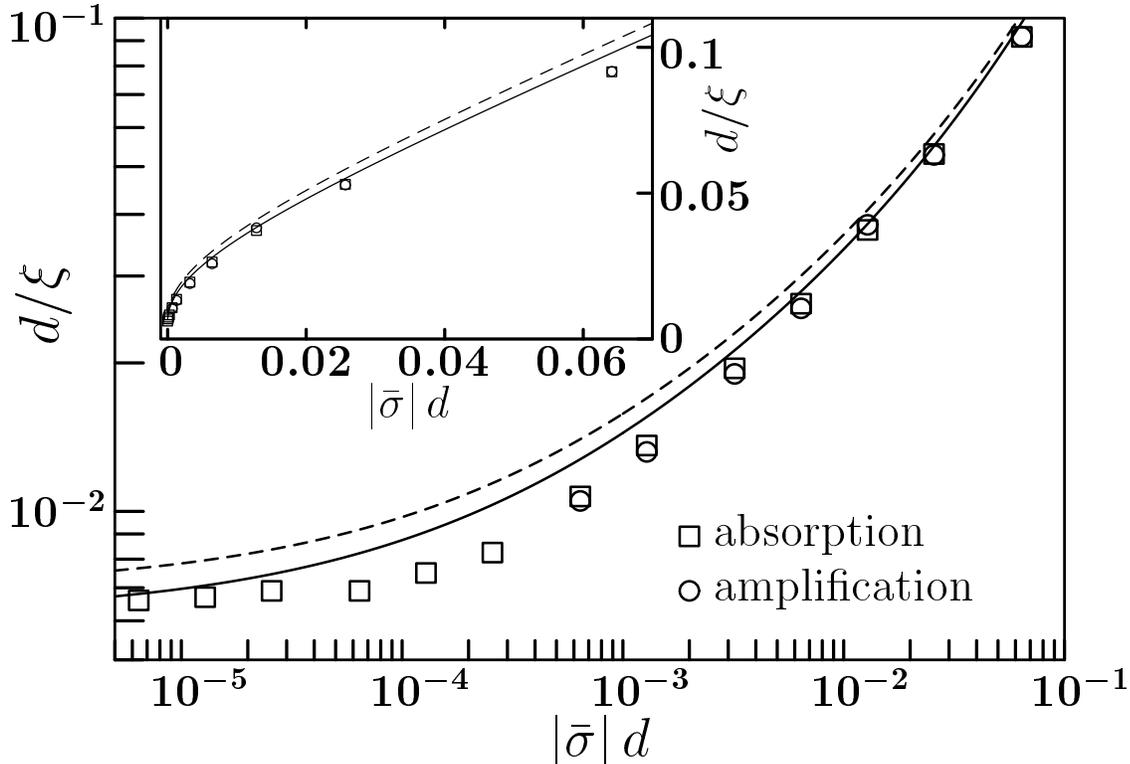}
\caption{
Localization length $\xi=-\lim_{L\to\infty}L^{-1}\langle\ln T\rangle$
of a disordered waveguide ($N=10$) versus the modal average
$\bar\sigma$ of the inverse absorption or amplification length. Data
points are a numerical solution of the discretized
(lattice constant $d$) two-dimensional
Helmholtz equation for the case of absorption (squares) or
amplification
(circles). The curves are the analytical prediction
(\protect\ref{ximax})
in the case $\beta=1$ (unbroken time-reversal symmetry) for
$l=29.6\,d$ [solid curve, determined from Eq.~(\protect\ref{xinuldef})]
and for $l=26.1\,d$ [dashed curve, determined from
Eq.~(\protect\ref{ximetaldef})].
The inset
shows the same data on a linear, rather than logarithmic, scale.}
\label{fig1}\end{figure}

The parameter $\bar\sigma$ is obtained from the analytical solution of
the discretized Helmholtz equation in the absence of disorder
($\delta\varepsilon=0$). The complex longitudinal wavenumber $k_{n}$ of
transverse mode $n$ then satisfies the dispersion relation
\begin{equation}
\cos(k_{n}d)+\cos(n\pi d/W)=2-\case{1}{2}(kd)^{2}(1+{\rm
i}\varepsilon''), \label{dispersion}
\end{equation}
which determines $\bar\sigma$ according to $\bar\sigma=-2N^{-1}\,{\rm
Im}\,\sum_{n}k_{n}$.
Simulations with $\varepsilon''=0$ were used to obtain $l$, either
from the large-$L$ relation\cite{Dor82}
\begin{equation}
  -\lim_{L\to\infty} L^{-1} \langle\ln T\rangle =
  [\case12(N+1)l]^{-1},
  \label{xinuldef}
\end{equation}
or from the large-$N$ relation \cite{Mel91}
\begin{equation}
\lim_{N\to\infty}\langle T\rangle=(1+L/l)^{-1}.
\label{ximetaldef}\end{equation}
The parameters chosen were $W=25\,d$, $k=1.22\,d^{-1}$, corresponding
to $N=10$, $l=29.6\,d$
from Eq.~(\ref{xinuldef}) and $l=26.1\,d$ from Eq.~(\ref{ximetaldef}).
 The localization length was computed
as a function of $\sigma$ from the $L$-dependence of $\ln T$ up to
$40\,l$, averaged over 150 realizations of the disorder. Results are
shown
in Fig.~\ref{fig1}. The localization length is the same
for absorption and amplification, within the numerical accuracy.
Comparison with the analytical result (\ref{ximax}) for $\beta=1$
is plotted for the two values of the mean free path. The agreement
is quite reasonable, given the approximate nature of Eq.~(\ref{ximax})
in the regime $|\gamma|N^2\simeq1$ (corresponding to
$|\bar\sigma|d\simeq10^{-4}$).

In conclusion, we have shown how absorption or amplification can be
incorporated into the Dorokhov-Mello-Pereyra-Kumar
equation for the transmission through a multi-mode
waveguide. The technical difficulty of the multi-mode case is
that the Fokker-Planck equation for the transmission eigenvalues ${\cal
T}_n$
depends not just on the transmission and reflection eigenvalues ${\cal
T}_n$,
${\cal R}_n$, but also on the eigenvectors of the matrices $tt^{\dagger}$
and $rr^{\dagger}$.
We could compute the localization length in the two regimes,
$|\gamma|\gg1/N^2$ and $|\gamma|\ll1/N^2$, and have given an
interpolation
formula for the intermediate regime. An exact solution for all
$\gamma$ remains an unsolved problem.

\acknowledgements
We acknowledge useful discussions with P.~W.~Brouwer and K.~M.~Frahm.
This research was supported in part by the ``Nederlandse organisatie
voor Wetenschappelijk Onderzoek'' (NWO) and by the ``Stichting voor
Fundamenteel Onderzoek der Materie'' (FOM).

\end{document}